\documentclass[aps,prb,twocolumn,floatfix]{revtex4-1}
\usepackage{color,graphicx}
\usepackage{amsmath,amssymb}
\input{amssym.def}

\begin{document}

\title{Jacobi Elliptic Functions and the\\Complete Solution to the Bead on the Hoop Problem}
\author{Thomas E.~Baker}
\author{Andreas Bill}\email{Andreas.Bill$@$csulb.edu; author to whom correspondence should be addressed.}
\affiliation{California State University Long Beach, Department of Physics \& Astronomy, 1250 Bellflower Blvd., Long Beach, CA 90840}
\date{submitted June 13, 2011}

\begin{abstract}
Jacobi elliptic functions are flexible functions that appear in a variety of problems in physics and engineering. We introduce  and describe important features of these functions and present a physical example from classical mechanics where they appear: a bead on a spinning hoop. We determine the complete analytical solution for the motion of a bead on the driven hoop for arbitrary initial conditions and parameter values.
\end{abstract}

\maketitle

\section{Introduction}\label{s:intro}

In 1788, James Watt introduced a mechanical device into steam engines that moderated the flow of steam and controlled the speed of the engine.\cite{Hart49,Dickinson67}  This device, known as the governor, is composed of two solid pendula (weights) fixed to a common vertical axis.  The power of the steam engine rotates the axis and lifts the weights to a height determined by the angular speed. The behavior of the weights is nowadays presented to students as an example of Lagrangian mechanics as a mass constrained to move frictionlessly on the surface of a sphere with constant azimuthal frequency or a bead moving on a rotating hoop.\cite{Goldstein02,Taylor05}

Solving the equation of motion of the bead on a rotating hoop is generally avoided as it does not have a straightforward solution.  The problem is usually limited to the determination of the stationary position of the mass for a given angular frequency of the hoop, although some papers treat the complete problem numerically or within some approximations.\cite{Whittaker37,Bowman53,Lima10}  Here, we present the full analytical solution for the motion of the bead on the hoop which requires the use of Jacobi elliptic functions (JEFs).

There are a number of physics problems where JEFs provide an excellent approximation if not a complete description of the system.\cite{Erdos00,Fu01} These special functions are also amply used in the conformal mapping of engineering problems.\cite{Bowman53,Kinnersley} JEFs have some interesting features. For example, trigonometric and hyperbolic functions are special cases of Jacobi elliptic functions. If considered in the complex plane, JEFs are doubly periodic, a feature appearing, for example, in bifurcation theory of chaotic systems.\cite{Ochoa06}

Despite the diversity of physical problems where JEFs appear, these functions are generally not part of the physics curriculum. This motivates us to offer a concise and pedagogical exposition of these functions. We then demonstrate how these special functions appear in the example of the bead on the hoop.

\section{Introducing Jacobi Elliptic functions}\label{s:Jacobi}

In this section, we introduce Jacobi elliptic functions through elliptic geometry and integral inversion,\cite{Bowman53,Baker1906,Armitage06,Prasolov97,Chandrasekharan85,Neville71} and discuss the geometrical interpretation of their arguments.
 
\subsection{Definitions and relations between Jacobi elliptic functions}\label{ss:JacobiDef}
 
We start by redefining the basic trigonometric functions sine and cosine in terms of the functional inverse of specific integrals. Usually, these functions are introduced using circular geometry. That is, for a given point $(x,y)$ on a circle of radius $r$, $\sin\theta = y/r$ and $\cos\theta = x/r$. An alternative way to define these functions is to realize that each {\it inverse} trigonometric function is a solution of a definite integral. For example, simple trigonometric substitutions lead to 
\begin{eqnarray}\label{integarcsinarctan}
\int_0^y \frac{dt}{\sqrt{1-t^2}} = \arcsin y, \quad \int_0^y \frac{dt}{1+t^2} = \arctan y,
\end{eqnarray}
with the conditions that $-1\leq y\leq 1$ and $\sqrt{1-t^2}\geq 0$.
We can  reinterpret these relations by stating that they {\it define} the inverse trigonometric functions. Following Abel we obtain trigonometric functions by inverting these integrals.\cite{Prasolov97} For example, defining the argument $\theta$ as
\begin{equation}\label{arcsinIntegral}
\theta(y) = \int_0^y \frac{dt}{\sqrt{1-t^2}} = \arcsin y,
\end{equation}
we obtain the trigonometric sine function $\sin \theta = y$. With the Pythagorean relation, we can define the cosine function and all others follow. This alternative definition implies that the integrals $\theta(y)$ must in particular display the parity and periodicity of the corresponding trigonometric functions $y(\theta)$.

Consider now the case of elliptic geometry. Gauss, Legendre, Abel, Jacobi and Weierstrass studied in depth properties of elliptic functions and integrals. For the purpose of this paper we note only the result, proved by Legendre,\cite{Baker1906} that any integrated expression containing a third or fourth degree polynomial in the denominator of a fraction can be reduced to a linear combination of the following elliptic integrals of the first, second, and third kind,
\begin{eqnarray}
\label{FirstKind}
F\left(y,k\right) & = & \int_0^y \dfrac{dt}{\sqrt{\left(1-t^{2}\right)\left(1-k^{2}t^{2}\right)}}\nonumber\\
F\left(\phi,k\right) &=& \int_0^\phi\dfrac{d\phi'}{\sqrt{1-k^2\sin^{2}\phi'}}\\
\label{SecondKind}
E\left(y,k\right) & = &  \int_0^y \sqrt{\dfrac{1-k^{2}t^{2}}{1-t^{2}}}dt\nonumber\\
E\left(\phi,k\right) &=& \int_0^{\phi} \sqrt{1-k^2\sin^2\phi'} \,d\phi'\\
\label{ThirdKind}
\Pi\left(y,k,n\right) & = &  \int_0^y \dfrac{dt}{\left(1-n^{2}t^{2}\right)\sqrt{\left(1-t^{2}\right)\left(1-k^{2}t^{2}\right)}}\nonumber\\
\Pi\left(\phi,k,n\right) &=& \int_0^\phi \frac{d\phi'}{\left(1-n^2\,\sin^2\phi'\right)\sqrt{1-k^2\,\sin^2\phi'}},
\end{eqnarray}
where $y=\sin\phi$, $t=\sin\phi'$, $k\in (-1,1)$ is the modulus and $n\in \mathbb{R}$ the characteristic. We assume all square roots to be positive. In Eqs.~\eqref{FirstKind}--\eqref{ThirdKind}, the first expression (in $t$) is known as the Jacobi form whereas the second (in $\phi'$) is Legendre's form. When $\phi = \pi/2$, written $F\left(\tfrac{\pi}{2},k\right)\equiv K\left(k\right)$, the integrals are said to be complete and otherwise incomplete.  We will use the first two elliptic integrals exclusively in what follows.

Because trigonometric functions are sometimes more useful than their inverses, it may be worthwhile to invert the elliptic integrals as well.\cite{Bowman53,Prasolov97} In this process, Abel and Jacobi followed the work of Legendre and introduced what are now known as the Jacobi elliptic functions. These functions result from the inversion of the elliptic integral of the first kind. Introducing the Jacobi notation $u(\phi) \equiv F(\phi,k)$ for this integral, we can formally invert the Legendre form of the integral, Eq.~\eqref{FirstKind}, to obtain $\phi(u) = \mathrm{am}\, u$, the amplitude of $u$. In the Jacobi form, $u(y) = F(y,k)$, the inversion of Eq.~\eqref{FirstKind} leads one to define the sine-amplitude Jacobi elliptic function written in Glaisher's notation ($y\in [-1,1]$, $\phi\in [-\pi/2,\pi/2]$)
\begin{equation}\label{snuk}
\mathrm{sn}(u,k) \equiv y = \sin\phi = \sin\left(\mathrm{am}\, u\right).
\end{equation}
Note that the $k$-dependence is often implicit in the literature, resulting in the notation $\mathrm{sn}\,u$ instead of $\mathrm{sn}\left(u,k\right)$. The last two equalities in Eq.~\eqref{snuk} relate the Legendre and Jacobi forms of the elliptic integrals and justify the name given to this Jacobi elliptic function. Because Eq.~\eqref{FirstKind} is odd, $u(y) = F(y,k) = \mathrm{sn}^{-1}(y,k)$ is odd in $y$ as well.  That is, the sine-amplitude is an odd function of $u$.

It is natural to introduce two further Jacobi elliptic functions. Taking the cosine of the amplitude of $u$, we define the cosine-amplitude
\begin{equation}\label{cnuk}
\mathrm{cn}(u,k) \equiv \cos\phi = \cos\left(\mathrm{am}\,u\right).
\end{equation}
An alternative way to define this function is to write $\mathrm{cn}\,u \equiv \sqrt{1-\mathrm{sn}^2u}$ since, from Eq.~\eqref{snuk}, $y\in[-1,1]$.

\begin{figure}
\begin{centering}
\includegraphics[scale=0.35]{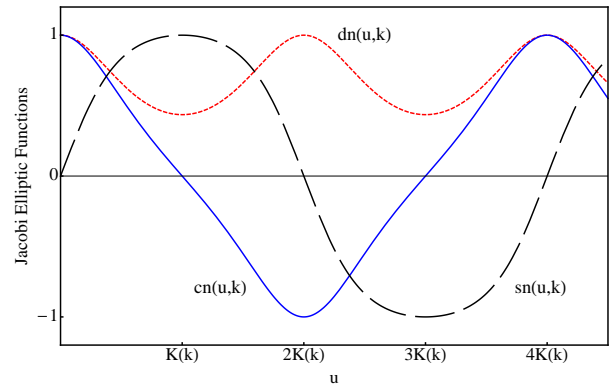}
\par\end{centering}
\caption{\label{fig:JEFu} (Color online)
Representative graphs of the Jacobi elliptic functions sn$(u)$, cn$(u)$, and dn$(u)$ at fixed value of the modulus $k=0.9$. In the bead and the hoop problem, that argument is a function of time.}
\end{figure}

Finally, in the Legendre form all elliptic integrals in Eqs.~\eqref{FirstKind}--\eqref{ThirdKind} contain $\Delta(\phi) \equiv \sqrt{1 - k^2\,\sin\phi}$. This leads one to define the delta-amplitude JEF that can also be written as a derivative of the amplitude\cite{Neville71}
\begin{equation}\label{dnuk}
\mathrm{dn}(u,k) \equiv \sqrt{1-k^2\mathrm{sn}^2u} = \frac{d(\mathrm{am}u)}{du}.
\end{equation}
Equation~\eqref{FirstKind} can also be used to obtain the last equality.
Figure~\ref{fig:JEFu} displays the characteristic behavior of the sine-, cosine- and delta-amplitude functions.

From the definitions above immediately follow the identities
\begin{eqnarray}
\mathrm{sn}^2u + \mathrm{cn}^2u &=& 1\label{JEFconditions1},\\
\mathrm{dn}^2u + k^2\,\mathrm{sn}^2u &=& 1\label{JEFconditions2},\\
\mathrm{cn}^2u + (1-k^2)\,\mathrm{sn}^2u &=& \mathrm{dn}^2u.\label{JEFconditions3}
\end{eqnarray}

For limiting values of $k$, the Jacobi elliptic functions reduce to trigonometric and hyperbolic functions
\begin{eqnarray}\label{Jacobispecialcase}
k&=&0:\,\, \mathrm{sn}u = \sin\,u, \,\, \mathrm{cn}\,u=\cos\,u, \,\, \mathrm{dn}u = 1,\\
k&=&1:\,\, \mathrm{sn}u = \tanh\,u, \,\, \mathrm{cn}\,u = \mathrm{dn}u = \mathrm{sech}\, u.
\end{eqnarray}
Jacobi elliptic functions thus include trigonometric and hyperbolic functions as special cases. However, JEFs are more than a simple generalization of elementary functions as can be seen when studying the properties of these functions in the complex plane (see  Appendix \ref{a:JEFrelations}).

Finally, we mention that from the definitions \eqref{snuk}--\eqref{dnuk}, one can define quotients of JEFs. One defines $\mathrm{sc}\,u = \mathrm{sn}\,u/\mathrm{cn}\,u$ which carries the name tangent-amplitude $\mathrm{sc}\,u\equiv\mathrm{tn}\,u$. Other quotients are, for example, $\mathrm{nd}\,u = 1/\mathrm{dn}\,u$, $\mathrm{sd}\,u = \mathrm{sn}\,u/\mathrm{dn}\,u$, etc.  Further properties of JEFs that are used in the treatment of the bead on the hoop are summarized in  Appendix~\ref{a:JEFrelations}.

\subsection{Geometric interpretation of the arguments $(u,k)$}\label{ss:JacobiGeometry}
 
In the case of trigonometric functions, such as $y=\sin \theta$, the argument $\theta$ has an unequivocal geometric interpretation in circular geometry; it is the angle between the $x$-axis and the radius at a specific point $(x,y)$ on the circle. The question arises whether in a JEF the argument $u$ and the modulus $k$ have a geometric meaning as well. We first caution that the argument of a special function (Bessel, Hypergeometric, etc.) does not generally have such unequivocal  physical or geometric meaning. For JEFs, the geometric definition of the argument $k$ is unique but not for the argument $u$, an important point not discussed explicitly in the literature.

\begin{figure}[h]
\begin{centering}
\includegraphics[scale=0.375]{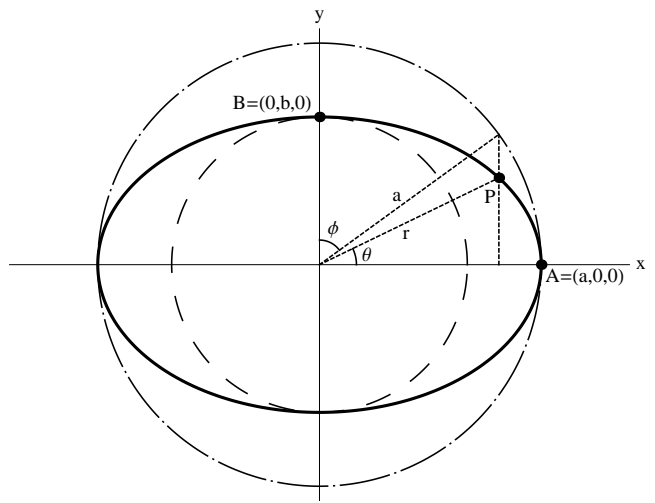}
\par\end{centering}
\caption{\label{fig:ellipse} Representations of points and angles on the ellipse that appear in the definitions and relations involving Jacobi elliptic functions. The figure shows the polar angle $\theta$ and the complementary of the eccentricity angle $\phi$. The smaller circle of radius $b$ (dashed line) is inscribed in the ellipse and has common points at $\pm B=\left(0,\pm b,0\right)$. The ellipse is also inscribed in the larger circle with radius $a$ (dash-dotted line) which has two points in common $\pm A=\left(\pm a,0,0\right)$.}
\end{figure}

The modulus $k$ is given in elliptic geometry by $k=\epsilon$, the eccentricity of the ellipse $\epsilon^2 = 1 - b^2/a^2$, where $a>b$ are the semi-axes (see Fig.~\ref{fig:ellipse})
\begin{equation}\label{ellipse}
\frac{x^2}{a^2} + \frac{y^2}{b^2} = 1.
\end{equation}
The equivalence between $k$ and $\epsilon$ is found by realizing that the arc length $s_{BP}$ on the ellipse is expressed in terms of the elliptic integral of the second kind. Using the parametrization $x=a\,\sin\phi$, $y=b\,\cos\phi$, the infinitesimal arc length along the ellipse is
\begin{eqnarray}
\label{dsellipse}
ds & = &\sqrt{{dx}^{2}+{dy}^{2}} = \sqrt{a^2\cos^2\phi +b^2\sin^2\phi}\,\,d\phi \nonumber \\
& = &a\sqrt{1-\epsilon^2\sin^2\phi}\,\,d\phi.
\end{eqnarray}
Using Eq.~\eqref{SecondKind}, the arc length $s_{BP}$ of the ellipse from $B$ to $P$ can be written
\begin{eqnarray}\label{SecondKindArc}
s_{BP} = a \int_0^{\phi} \sqrt{1-\epsilon^2\sin^2\phi'} \,d\phi' = a\,E(\phi,k=\epsilon).
\end{eqnarray}
This relation between the arc length of the ellipse and the function $E(\phi,k)$ is the reason why Fagnano (cited in Ref.~\onlinecite{Prasolov97}) called $F$, $E$, and $\Pi$ elliptic integrals and provides a direct interpretation of the modulus of the JEFs.

The geometric interpretation of JEFs' argument $u$ is less direct and not unique. Fagnano discovered that for $k^{-1}=\sqrt{2}$, the elliptic integral of the first kind describes the arc length on the lemniscate defined by the relation $r^2 = \cos 2\theta$ (using the same definitions of $r$ and $\theta$ as in Fig.~\ref{fig:ellipse}).\cite{Prasolov97,Armitage06} Serret, building on the work of Fagnano and Legendre, found a set of curves on which $F(\phi,k)$ at arbitrary value of $k\in(-1,1)$ is an arc length (see, {\it e.g.} Ref.~\onlinecite{Prasolov97}).  There is also a counterpart in three dimensions: Seiffert's spirals defined on a sphere. The arc length of the spiral is given by the first elliptic integral.\cite{Erdos00} In these examples the argument $u$ of the JEF is the length of the Serret curve or the Seiffert spiral.

Additional geometric interpretations of the argument $u$ can be found from trigonometry as it applies to an ellipse using geodesy and elliptic geometry. This is discussed in detail in Refs.~\onlinecite{Bowman53} and \onlinecite{bakerMS12}. All of these interpretations relate $u$ to a length of some geometric form related to an ellipse. It is important, however, to realize that the angles used to define $u$ are different for each physical situation. This leads to the important insight that the argument $u$ has no unique geometric interpretation beyond being an arc length and only has a physical interpretation in some special cases.\cite{Bowman53,Armitage06,bakerMS12}

\section{The Bead on the Hoop}\label{s:beadhoop}

\begin{figure}[h]
\begin{centering}
\includegraphics[scale=0.2]{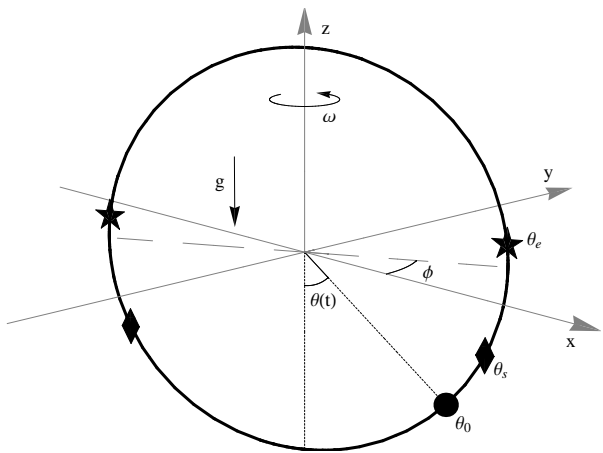}
\par\end{centering}
\caption{\label{fig:BeadandHoop}Schematic of a bead on a hoop rotating about the vertical $z$-axis at a constant azimuthal frequency $\omega=\dot{\phi}$. The diamond at $\theta = \pm\theta_s$ locates the global minimum of the effective potential at the commonly named stationary point, whereas the star at $\theta = \pm\theta_e$ locates the so-called ejection point (see text and Fig.~\ref{fig:Veff}). The positions of $\theta_s$ and $\theta_e$ are obtained using $\omega=3\omega_c/2$. The motion of the bead is described by the polar angle $\theta(t)$ measured from the bottom of the hoop.  }
\end{figure}

In this section we present an application of Jacobi elliptic functions to a well-known problem.  Consider a bead of mass $m$ that slides without friction on a hoop of radius $R$ rotating with constant angular velocity $\omega=\dot{\phi}$ (see Fig.~\ref{fig:BeadandHoop}).  In spherical coordinates $(r,\theta,\phi)$ the Lagrangian $\mathcal{L} = T -  V$ for this system is
\begin{equation}\label{eq:Lagrangian}
\mathcal{L}\left(\theta,\dot{\theta},t\right)  = \frac{1}{2}I_0 \left(\dot{\theta}^{2}+\omega^{2}\sin^{2}\theta\right)-V_0\left(1-\cos\theta\right),
 \end{equation}
where $I_0 = mR^2$ and $V_0 = mgR$. The zero point of the potential energy is located at the bottom of the hoop. The angles $\theta$ and $\phi$ should not be confused with the angles of same name in Sec.~\ref{s:Jacobi}. The polar angle $\theta$ is measured from the bottom of the hoop for consistency with the convention usually used in the limiting case of the simple non-linear pendulum.

There are three variables $(r,\theta,\phi)$ and two constraints $r=R$ and $\dot\phi=\omega$, implying that there is only one generalized coordinate, $\theta$. That is, the motion of the bead is described by the time-dependent function $\theta\left(t\right)$.

\begin{figure}
\begin{centering}
\includegraphics[scale=0.39]{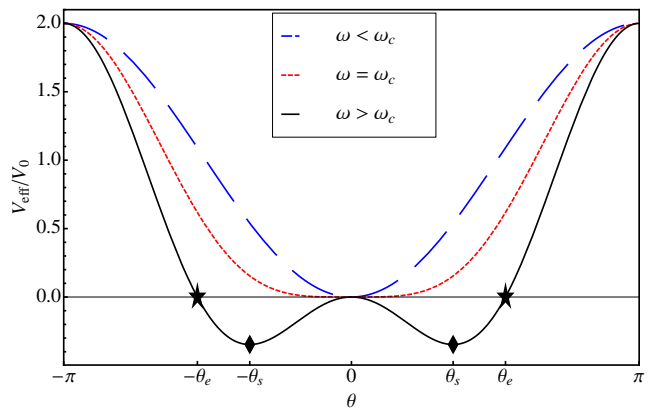}
\par\end{centering}
\caption{\label{fig:Veff} (Color online) The effective potential \eqref{eq:Veff} as a function of $\theta$. The dashed line is for $\omega<\omega_c$ (for the present plot we chose $\omega= \omega_c/5$), the dotted line for $\omega = \omega_c$, and the solid line is for $\omega>\omega_c$ (here $\omega = 3\omega_c/2$). The diamond at $\pm\theta_s$ and star at $\pm\theta_e$ are indicated in Fig.~\ref{fig:BeadandHoop} for the present set of parameters. The effective potential energy at $\theta_e$ and $\theta=0$ is always zero. Note that $V_{\mathrm{eff}}^\mathrm{max} = V_\mathrm{eff}(\pi) = 2V_0$.}
\end{figure}

The discussion of the physical content of the following calculations is greatly aided by the introduction of an effective potential $V_\mathrm{eff}$. Since $\theta$ is the only degree of freedom, we combine the azimuthal kinetic energy with the gravitational potential energy and rewrite Eq.~\eqref{eq:Lagrangian} as $\mathcal{L}\left(\theta,\dot{\theta},t\right) = T_{\theta}\left(\dot{\theta}\right) -  V_\mathrm{eff}(\theta)$, with
\begin{eqnarray}\label{eq:Veff}
\frac{V_\mathrm{eff}\left(\theta\right)}{V_0} &=& - \frac{1}{2} \left(\frac{\omega^{2}}{\omega_c^2}\right) \sin^{2}\theta + \left(1-\cos\theta\right)
\end{eqnarray}
and $\omega_c^2 = V_0/I_0 = g/R$.    As shown in Fig.~\ref{fig:Veff}, the qualitative shape of the effective potential depends on the azimuthal frequency $\omega$. One distinguishes two regimes: $\omega \leq \omega_c$ when $\ddot\theta\leq 0$, and $\omega>\omega_c$ when $\ddot\theta>0$.  Since $\theta_0 = \pi$ is always a maximum of $V_\mathrm{eff}$, this position is unstable. On the other hand, $\theta_0=0$ is a minimum in the first regime but a local maximum in the second regime where two new stable minima appear at $\pm\theta_s$.  For energies near the minima at $\pm\theta_s$ , the bead will oscillate about $\theta_s$, remaining exclusively on one side of the hoop.  The stable equilibrium point $\theta_s$ can be obtained by setting the derivative of the effective potential equal to zero. We find
\begin{equation}\label{thetas}
\theta_s(\omega) =  \arccos \left(\frac{\omega_c^2}{\omega^2}\right),
\end{equation}
which is valid in the supercritical regime ($\omega>\omega_c$).

The differential equation governing $\theta(t)$ is obtained from the Euler-Lagrange equation associated with Eq.~\eqref{eq:Lagrangian}
\begin{equation}\label{eq:EqofMotion}
\ddot{\theta}=\frac{\omega^{2}}{2}\sin2\theta-\omega_c^2\sin\theta,
\end{equation}
and introduce the initial conditions $\theta(t=0) = \theta_0$ and $\dot{\theta}(t=0)=\dot{\theta}_0$. Without loss of generality, we make use of the symmetry of the problem and define the initial angle $\theta_{0}$ only on one half of the hoop $\theta_{0}\in\left[0,\pi\right]$.  Though nonlinear, this differential equation can be solved analytically. The presence of $\sin\theta$ and $\sin2\theta$ leads to solutions involving JEFs described in Sec.~\ref{s:Jacobi}.

\subsection{General Solution of the Euler-Lagrange Equation}\label{ss:EulerLagrange}

We begin by multiplying Eq.~\eqref{eq:EqofMotion} by $\dot\theta$ and integrating to obtain
\begin{equation}\label{eq:Firstintegration}
\frac{\dot\theta^2(t)}{\omega^2} + \cos^2\theta(t) - 2\cos\theta_s\,\cos\theta(t) = {\cal C},
\end{equation}
where the constant ${\cal C}$ is given by
\begin{equation}\label{eq:E}
{\cal C} \equiv \cos^{2}\theta_{0}-2\cos\theta_{s}\cos\theta_{0}+\left(\frac{\dot{\theta}_0}{\omega}\right)^2,
\end{equation}
and $\theta_s$ is given in Eq.~\eqref{thetas}. Note that Eq.~\eqref{eq:Firstintegration} is not valid for $\omega=0$; this situation must be considered separately (see Sec.~\ref{ss:pendulum}).  It is worth mentioning that $\theta_s$ is nonzero only for $\omega > \omega_c$ and takes values in the interval $\theta_s\in [0,\pi/2]$. For $\omega < \omega_c$, only $\theta_0 = 0$ is a stable stationary solution.  It is then natural to extend the definition of $\theta_s$ by imposing $\theta_s=0$ for the case where $\omega<\omega_c$.

Because ${\cal C}$ is a constant, the left hand side of Eq.~\eqref{eq:Firstintegration} is a constant of motion of the system. The expression for ${\cal C}$ is of particular interest because it allows us to reduce arbitrary initial conditions to ones with zero initial speed. Consider arbitrary initial conditions $(\tilde{\theta}_0,\dot{\tilde{\theta}}_0)$ for which $\dot{\tilde{\theta}}_0 \neq 0$ and $E_\mathrm{tot}<V_\mathrm{eff}^\mathrm{max}$. From Eq.~\eqref{eq:Firstintegration} we can find a time $t_0$ and new initial conditions $(\theta_0,\dot{\theta}_0)$ with zero initial angular speed  $\dot{\theta}_0 = 0$  such that $\theta(t_0) = \tilde{\theta}_0$ and $\dot{\theta}(t_0) = \dot{\tilde{\theta}}_0$.
The new initial condition $(\theta_0,0)$ written in terms of the actual one $(\tilde{\theta}_0,\dot{\tilde{\theta}}_0)$ is
\begin{equation}\label{eq:RestingInitialAngle}
\cos\theta_{0}=\cos\theta_{s}\pm\sqrt{\left[\cos\tilde{\theta}_0 -\cos\theta_{s}\right]^{2}+ \frac{1}{\omega^{2}}\dot{\tilde{\theta}}_0^{2}}.
\end{equation}
Thus, arbitrary initial conditions with $\dot{\tilde{\theta}}_0 \neq 0$ can be written in terms of initial conditions with $\dot\theta_0 = 0$.
In the following, we limit our discussion to the motion of the bead on the hoop with zero initial speed unless $E_\mathrm{tot}>V_\mathrm{eff}^\mathrm{max}$. This case will be discussed later. 

We now use the Weierstrass substitution\cite{Spivak06}
\begin{equation}\label{tautransf}
\tau=\tan\left(\frac{\theta}{2}\right) =\sqrt{\frac{1-\cos\theta}{1+\cos\theta}},
\end{equation}
in Eq.~\eqref{eq:Firstintegration} and obtain 
\begin{equation}\label{eq:Integral}
\int_{\tau_0}^{\tau\left(t\right)}\frac{2d\tau'}{\sqrt{p_{-}\,\tau'^{4}+2\left({\cal C}+1\right)\tau'^{2}+p_{+}}}=\omega t
\end{equation}
through separation of variables. This type of equation is usually studied numerically  or in limiting cases only.\cite{Whittaker37,Lima10,Bowman53}  Here we extend the calculation and invert the integral to obtain $\theta(t)$.  The structure of the integral implies that the inverse is expressed in terms of JEFs defined in Sec.~\ref{s:Jacobi}.

For convenience, we define $\tau_0 = \tan\left(\theta_0/2\right)$  and the two constants
\begin{eqnarray}\label{eq:ppm}
p_{\pm} &\equiv& {\cal C}-1\pm2\cos\theta_{s}.
\end{eqnarray}
After factoring out $p_{-}$ from the denominator of the integrand in Eq.~\eqref{eq:Integral}, we write the remaining polynomial under the square root in the form
\begin{equation}\label{taugamma12}
\tau'^{4}+2\left(\frac{{\cal C}+1}{p_-}\right){\tau'}^{2}+\left(\frac{p_+}{p_-}\right) = \left({\tau'}^2 - \gamma_1^2\right)\left({\tau'}^2 - \gamma_2^2\right),
\end{equation}
where the roots are given by
\begin{equation}\label{gamma12}
\gamma_{1,2}^{2}\in
\left\{ a,b \right\}
\end{equation}
with
\begin{eqnarray}\label{abroots}
a \equiv  \frac{\textstyle 1-\cos\theta_{0}}{\textstyle 1+\cos\theta_{0}}= \tau_0^2,\quad
b \equiv \frac{\textstyle 1+\cos\theta_{0}-2\cos\theta_{s}}{\textstyle 1-\cos\theta_{0}+2\cos\theta_{s}}.
\end{eqnarray}
The root $a$ is always non-negative while $b$ can take any real value. As a result, the relative values of the two roots $a$ and $b$ divides the range $\theta_0\in[0,\pi]$ into one (for $\omega\leq\omega_c$) or two (for $\omega>\omega_c$) intervals. When $\omega>\omega_c$, the two intervals are divided by the point $b=0$, which occurs when the initial condition $\theta_0$ takes the value
\begin{equation}\label{thetae}
\theta_{e}(\omega)\equiv\arccos \left(2\cos\theta_{s}-1\right) = \arccos \left(2\frac{\omega_c^2}{\omega^2} -1 \right).
\end{equation}
For $\theta_0<\theta_e$ we have $b>0$ while for $\theta_0>\theta_e$ we have $b<0$.  The angle $\theta_e$ appears in a natural way in our mathematical analysis and is new in the context of the bead and the hoop problem.  This angle is represented with a star in Figs.~\ref{fig:BeadandHoop} and \ref{fig:Veff}.

We call $\theta_e$ the ``ejection angle" because for any $\theta_0<\theta_e$, the bead will remain on the initial half of the hoop forever and never reach the other side.  However, when $\theta_0>\theta_e$ the bead has sufficient energy to cross the relative maximum at $\theta=0$ and eject itself from the original half of the hoop where it was released.  We note that $\theta_e(\omega_c) = \theta_s(\omega_c) = 0$ which compels us to extend their definition by imposing $\theta_e = \theta_s = 0$ for $\omega\leq\omega_c$. The two angles $\theta_s$ and $\theta_e$ also satisfy the inequality $\theta_e \geq \theta_s$ and $\theta_e \in [0,\pi]$ where $\theta_e(\omega\to\infty) \to \pi$.   Curiously, we observe that $V_g\left(\theta_e\right)=2V_g\left(\theta_s\right)$ but are unaware if this is the result of some physical constraint.
Further considerations where the bead is placed directly on this point are addressed in Sec.~\ref{ss:separatrices}. 

The integral Eq.~\eqref{eq:Integral}, along with Eq.~\eqref{taugamma12}, can be related to the first elliptic integral [Eq.~\eqref{FirstKind}]. Inversion of the integral gives $\tau(t)$ and thus $\theta(t)$ using Eq.~\eqref{tautransf}. Which particular inverse JEF equals Eq.~\eqref{eq:Integral} depends on $\gamma_1$ and $\gamma_2$ (or $a$ and $b$).

For $\omega\leq \omega_c$, we have $\theta_s=\theta_e=0$ and $a=b\geq 0$ for $\theta_0 \in [0,\pi]$. From the tables in Ref. \onlinecite{Byrd71} with $\gamma_{1}^{2}=a$ and $\gamma_{2}^{2}=b$ in Eq.~\eqref{taugamma12}, we identify the cosine-amplitude. Transforming back to the variable $\theta = 2\arctan\tau$, we obtain
\begin{eqnarray}\label{cnSolution}
\tan\left[\frac{\theta(t)}{2}\right] = \sqrt{a}\, \mathrm{cn}\left[\frac{1}{2}\omega t\sqrt{\left|p_{-}\right|\left(a+\left|b\right|\right)},\sqrt{\frac{a}{a+\left|b\right|}}\right].\quad
\end{eqnarray}

On the other hand, when $\omega>\omega_c$ the range $\theta_0\in[0,\pi]$ is divided into three intervals $[0,\theta_s]$, $[\theta_s,\theta_e]$, and $[\theta_e,\pi]$, each of which leads to a solution involving a different JEF. For $\theta_{0}\in\left[0,\theta_{s}\right]$ the roots obey the inequality $b>a\geq 0$; for $\theta\in\left[\theta_s,\theta_e\right]$ we have $a>b\geq 0$; and for $\theta\in[\theta_e,\pi]$ we have $a\geq 0\geq b$. Using again the table of integrals in Ref.~\onlinecite{Byrd71} we set $\gamma_{1}^{2}=a$ and $\gamma_{2}^{2}=b$ for regions $\theta_{0}\in\left(0,\theta_{s}\right)$ and $\theta_{0}\in\left(\theta_{e},\pi\right)$, whereas $\gamma_{1}^{2}=b$ and $\gamma_{2}^{2}=a$ for $\theta_{0}\in\left(\theta_{s},\theta_{e}\right)$. This allows identifying the Jacobi elliptic functions nd, dn, and cn, respectively. The third interval $\theta_e\leq\theta\leq\pi$, turns out to lead to an expression that is identical to Eq.~\eqref{cnSolution} above. 

One can reduce the three cases for $\omega > \omega_c$ to one single expression for the entire interval $[0,\pi]$ by using properties of the JEFs derived in Sec.~\ref{ss:JacobiDef} and  Appendix \ref{a:JEFrelations}. The simplest form of the solution is found in terms of the delta-amplitude function, which naturally appears in the second interval. The cosine-amplitude function appearing in Eq.~\eqref{cnSolution} can be transformed into a delta-amplitude with Eq.~\eqref{JEFcorrespondence2}. Similarly, using the relation discussed after Eq.~\eqref{addsndn}, $k'\,\mathrm{nd}(u,k) = \mathrm{dn}(u+K,k)$,  it is possible to reduce the ${\rm nd}$ solution in the first interval to a delta-amplitude function. Careful consideration of the argument $(u,k)$ in the transformation leads to the final result
\begin{eqnarray}\label{eq:Solution}
\theta\left(t\right) &=&
2\arctan\left\{\sqrt{a}\, \mathrm{dn}\left[\frac{1}{2}\omega t\sqrt{\left|p_{-}\right|\,a},\sqrt{\frac{a-b}{a}}\right]\right\},\nonumber\\
&&\hspace*{30ex} \theta_{0}\in\left[0,\pi\right].
\end{eqnarray}
This is the exact solution of the equation of motion \eqref{eq:EqofMotion}, which fully describes the motion of a bead on a spinning hoop for arbitrary initial conditions $(\theta_0,\dot{\theta}_0)$, excepting the cases where $E_\mathrm{tot} \geq V_g(\pi)$ or $\omega=0$. Equation~\eqref{eq:Solution} simplifies to $\theta(t) = \theta_s$ for all times if $\theta_0=\theta_s$.  This solution improves or complements previous studies of different variants of the problem.\cite{Taylor05,Ochoa06,Lima10,Rousseaux09,Bowman53,Whittaker37} 

We note that the argument $u(t)$ of the JEF is a linear function of time and is proportional to $\phi = \omega t$ given by Eq.~\eqref{eq:Integral}. As pointed out in Sec.~\ref{s:Jacobi},  it is only in special cases that the argument has a useful physical interpretation, and the bead on the hoop is not one of them. In fact, the problem is slightly more difficult than what was presented in the previous section because the JEF was only identified after performing a Weierstrass substitution on the integral of Eq.~\eqref{eq:Integral}. The Weierstrass substitution can be shown to be a particular mapping of the position of the bead onto the $xy$-plane. The map is obtained from the intersection between the $xy$-plane and the line passing through the top of the hoop and through the bead. The argument $u(t)$ is related to the arc length $(R \sqrt{|p_-|\,a}/2)\,\phi$ on this stereographic projection.\cite{bakerMS12}

\subsection{Limiting and Special Cases}\label{ss:Special}

We discuss the special cases not considered in the previous section: the separatrices ($E=0,\,2mgR$), the free spinning of the bead around the hoop ($E>2mgR$), and the general solution for the pendulum ($\omega=0$). For all these cases the roots \eqref{abroots} take values $a,b\to\infty$ or 0 and the determination of $\theta(t)$ requires special care.

\subsubsection{Separatrices}\label{ss:separatrices}

\begin{figure}
\begin{centering}
\includegraphics[scale=0.5]{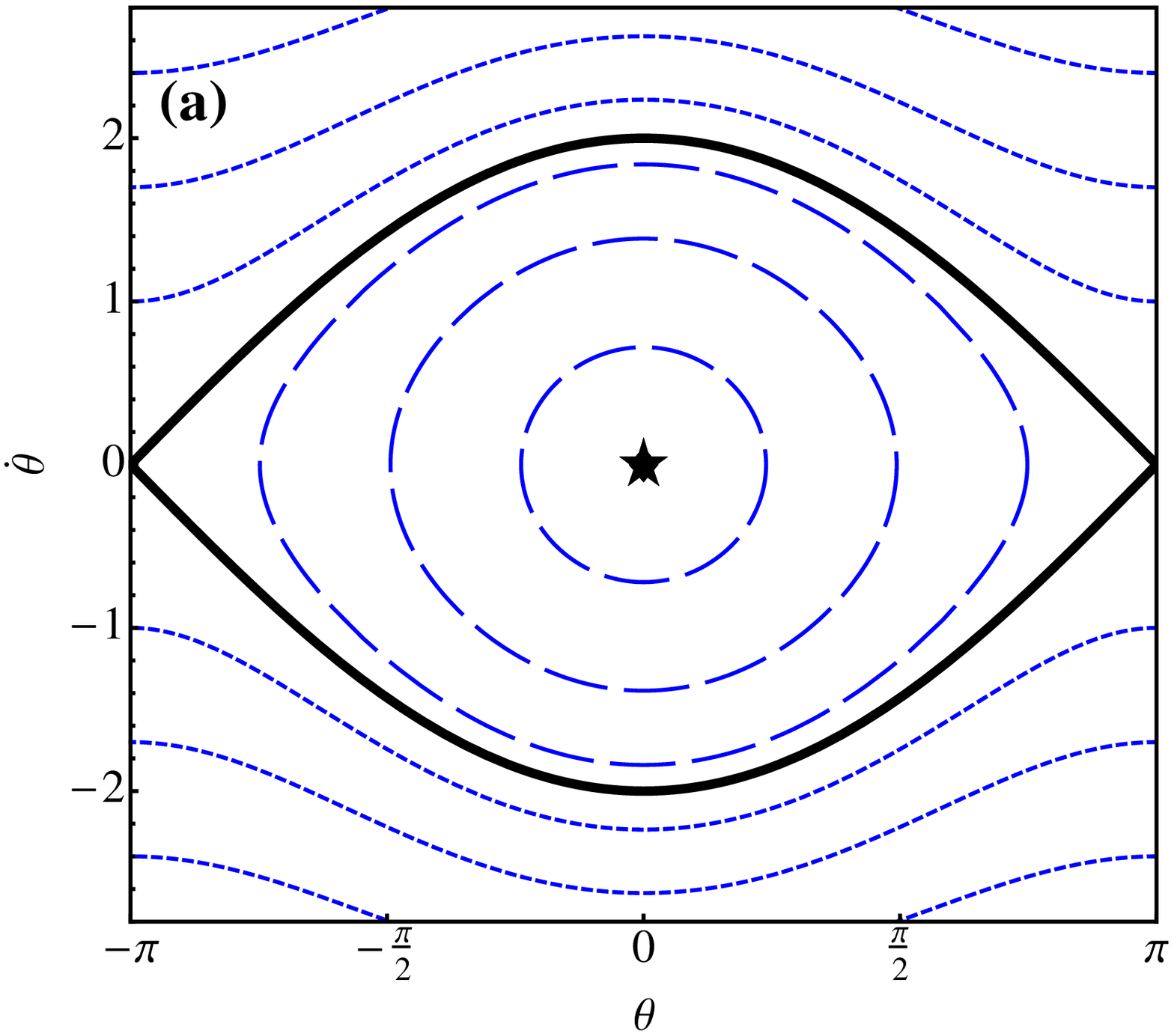}
\includegraphics[scale=0.5]{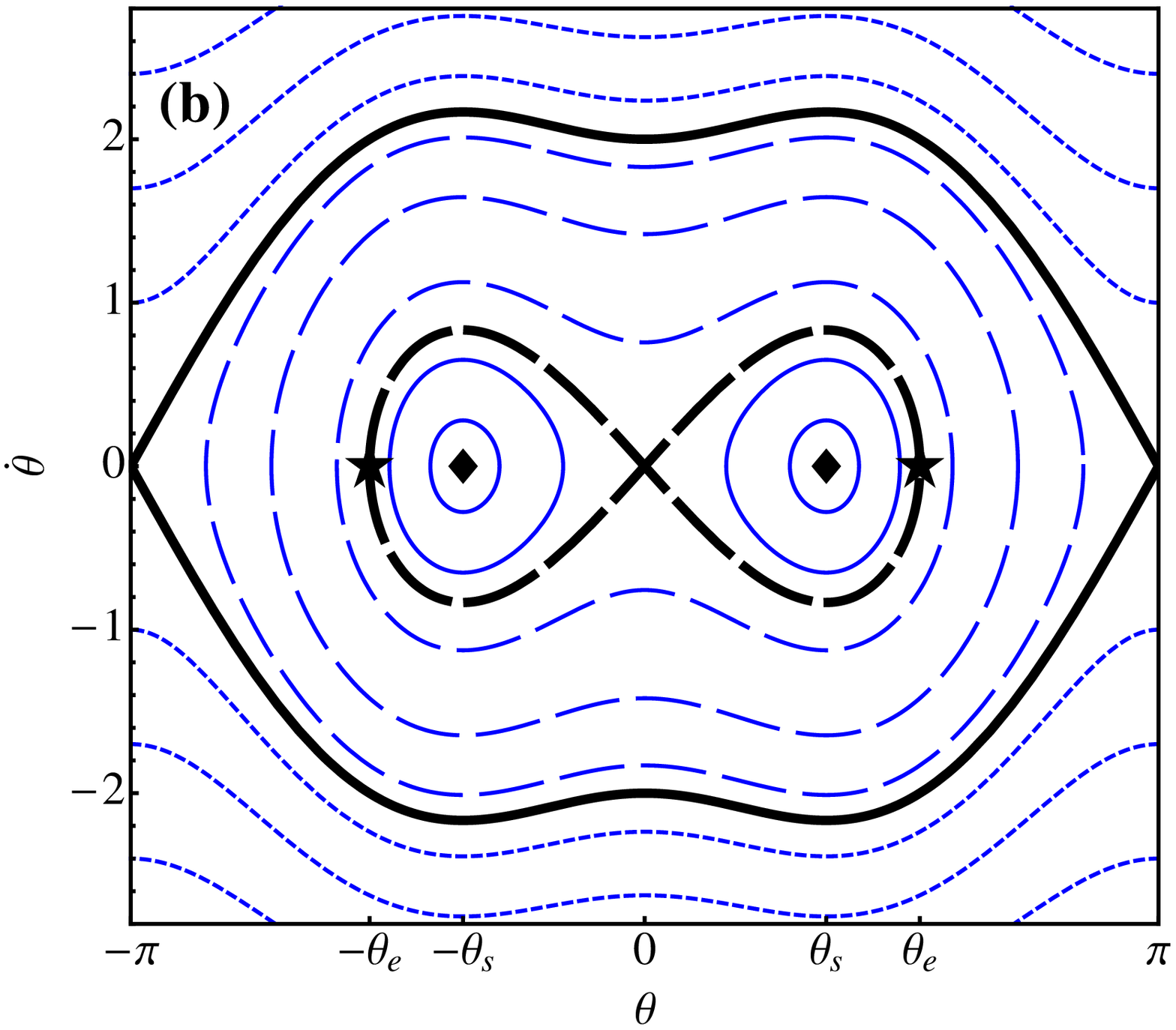}
\par\end{centering}
\caption{\label{fig:PhaseDiagrams} (Color online) Phase plots drawn from analytical expressions for (a) subcritical ($\omega \leq \omega_c$) and (b) supercritical ($\omega >\omega_c$) cases with the values given in Fig.~\ref{fig:Veff}. The stars ($\pm\theta_e$) and diamonds ($\pm\theta_s$) indicate the stable dynamical equilibrium points. The solid lines correspond to the periodic motion of the bead inside one of two potential wells. The dashed lines are motions of the bead across the two potential wells. The dotted lines correspond to the spinning of the bead around the hoop.  The two thick lines (solid and dashed) are the separatrices.}
\end{figure}

Figure~\ref{fig:PhaseDiagrams} shows phase plots for this problem.  The separatrices form the boundaries between qualitatively different motions of the system. They can be understood as delineating a phase change and typically involve functions which are asymptotic to some value. Below, we obtain simple analytical expressions for the separatrices that allow us to plot Fig.~\ref{fig:PhaseDiagrams} without the need of a numerical procedure applied to Eq.~\eqref{eq:Integral}.

As seen in Fig.~\ref{fig:PhaseDiagrams}b, there are two separatrices when $\omega>\omega_c$. The innermost separatrix (dashed thick line) passing through $(\theta_0,\dot\theta_0)=(0,0)$ separates the cases $\theta_0 <\theta_e$ from $\theta_e<\theta_0<\pi$. When $\theta_0<\theta_e$ the periodic motion is limited to either of the two wells of $V_\mathrm{eff}$ (see Fig.~\ref{fig:Veff}) and the bead oscillates exclusively on the half of the hoop where it started. By contrast, starting positions in the range $\theta_e<\theta_0<\pi$ involve the bead oscillating between the two sides of the hoop and moving through both wells. 

The expression $\dot\theta(\theta)$ of the inner separatrix is obtained from Eq.~\eqref{eq:Firstintegration} by noting that $\theta_0=\dot\theta_0=0$ is one point of the separatrix. At that point $\mathcal{C}=1-2\cos\theta_s$, which implies that any other point of the separatrix satisfies the equation
\begin{equation}\label{1stSeparatrix}
\dfrac{\dot{\theta}^2}{\omega^{2}} = \left(1-\cos\theta\right)\left(1 + \cos\theta - 2\cos\theta_{s}\right).
\end{equation}
The inner separatrix has the shape of a lemniscate as shown by the thick dashed line in Fig.~\ref{fig:PhaseDiagrams}b.

The motion on the separatrix is obtained from Eq.~\eqref{eq:Integral}. The result obtained for arbitrary initial condition on the separatrix is rather involved when $(\theta_0,\dot\theta_0) \neq (\theta_e,0)$ and will not be needed in the following. For the simpler case of $\theta_0 = \theta_e$, Eq.~\eqref{eq:Integral} or \eqref{eq:Solution} becomes 
\begin{eqnarray}\label{eq:Solutionthetae}
\theta(t) = 2\arctan\left[\sqrt{\frac{\omega^2}{\omega_c^2}-1}\,\,\,\mathrm{sech}\left(\omega_c\,t\sqrt{\frac{\omega^2}{\omega_c^2}-1}\right)\right],\quad
\end{eqnarray}
which is valid for $\omega>\omega_c$. We note that in this case the point $(\theta,\dot\theta)=(0,0)$ is an unstable equilibrium position. If the bead starts at any other initial condition (for example at $\theta_e$ and $\dot\theta_0=0$) the bead will need an infinite amount of time to reach the bottom of the hoop.

The second separatrix divides the phase plot between oscillations of the bead that reach both halves of the hoop, and the ``free" rotation of the bead around the hoop when $E_\mathrm{tot} > V_\mathrm{eff}^\mathrm{max}$. The equation of this separatrix can be obtained from Eq.~\eqref{eq:Firstintegration} by noting that $(\theta_0 = \pi,\dot\theta_0=0)$ is one point on the separatrix. At that point Eq.~\eqref{eq:E} reads ${\cal C} = 1+2\cos\theta_s$, which immediately implies from \eqref{eq:Firstintegration} that at any other point of the separatrix we have
\begin{equation}\label{eq:MaxDerivative}
\dfrac{\dot{\theta}^2}{\omega^{2}} = \left( 1+\cos\theta \right)\left( 1 - \cos\theta + 2\cos\theta_{s} \right).
\end{equation}
This separatrix is the thick solid line in Figs.~\ref{fig:PhaseDiagrams}a and \ref{fig:PhaseDiagrams}b. Any initial condition away from $(\theta_0,\dot{\theta}_0)=(\pi,0)$ on the separatrix will lead the bead to tend toward $\pi$ for $t\rightarrow\infty$. To obtain $\theta(t)$ for initial conditions lying on this second separatrix we integrate Eq.~\eqref{eq:Integral} to get
\begin{widetext}
\begin{eqnarray}\label{eq:Solutionthetapi}
\theta\left(t\right)&=&2\arctan\left(\sqrt{\frac{\cos\theta_{s}}{1+\cos\theta_{s}}}\sinh\left\{ \omega t\sqrt{1+\cos\theta_{s}}+\mathrm{arcsinh}\left[\tan\left(\frac{\theta_0}{2}\right)\,\sqrt{\frac{1+\cos\theta_{s}}{\cos\theta_{s}}}\right]\right\} \right).
\end{eqnarray}
\end{widetext}

The results of this subsection leads to the important observation that the separatrices are not described in terms of JEFs. Instead they depend solely on trigonometric and hyperbolic functions. A further observation is that the time evolution along the separatrices is not periodic but is asymptotic towards an unstable extremum.

\subsubsection{Free spinning}\label{ss:freespinning}
For initial derivatives $\dot\theta_0$ greater than the right hand side of Eq.~\eqref{eq:MaxDerivative}, we are outside the second separatrix in the phase plot and the bead is spinning freely on the hoop. Such motion is obtained, for example, when $\theta_0=\pi$ and $\dot\theta_0 \neq 0$ for which $E_{\mathrm{tot}}> V_\mathrm{eff}^\mathrm{max}$. Any other initial condition $(\theta_0,\dot\theta_0\neq 0)$ such that the total energy of the bead is larger than $V_\mathrm{eff}^\mathrm{max}$ can be mapped to this initial condition at $\theta_0=\pi$. To obtain the motion $\theta(t)$ of the bead in this special case it is, once more, easier to start with Eqs.~\eqref{eq:Firstintegration}--\eqref{eq:ppm}. For $(\theta_0=\pi,\dot\theta_0\neq 0)$ we have from Eq.~\eqref{eq:ppm} $p_- = \dot\theta_0^2/\omega^2$, $p_+ = 4\cos\theta_s + p_-$ and $\tau_0\to\infty$. The solution is written in terms of the tangent-amplitude function $\mathrm{tn}=\mathrm{sn}/\mathrm{cn}$. The integration of Eq.~\eqref{taugamma12} leads to
\begin{equation}\label{eq:FreeMotion}
\theta\left(t\right)=2\arctan\left[\frac{\left|\gamma_{-}\right|}{\mathrm{tn}\left(-\frac{1}{2} \left|\gamma_{-}\dot\theta_0\right|\,t,\sqrt{\frac{\left|\gamma_{-}^{2}\right|-\left|\gamma_{+}^{2}\right|}{\left|\gamma_{-}^{2}\right|}}\right)}\right],
\end{equation}
but with the roots $\gamma_{1/2} = \gamma_{+/-}$ given by
\begin{eqnarray}
\gamma_{\pm}^{2} = -2\left[\Gamma+\frac{1}{2}\right]
\pm 2 \sqrt{\Gamma^{2}+\frac{\omega^2}{2\dot\theta_0}\left(1-\cos\theta_s\right)},
\end{eqnarray}
with $\Gamma= (\omega^2 /\dot\theta_0^2)\left(1+\cos\theta_s\right)$. The oscillation of the bead corresponds to the open  dotted lines outside the second separatrix in Fig.~\ref{fig:PhaseDiagrams}b (and outside the separatrix of Fig.~\ref{fig:PhaseDiagrams}a for $\omega\leq\omega_c$).

For $\left|\dot{\theta}_0\right|\rightarrow\infty$, $\gamma_\pm=-1$ and Eq.~\eqref{eq:Integral} reduces to the form of the arctangent function given in Eq.~\eqref{integarcsinarctan}. This implies that as $\dot{\theta}_0$  increases without bound, our solution approaches the form $\theta(t)=\left|\dot{\theta}_0\right|t$.  We recognize this as characteristic of a free-particle solution.

\subsubsection{General solution for pendulum ($\omega = 0$)}\label{ss:pendulum}
The pendulum is described by Eq.~\eqref{eq:EqofMotion} for $\omega=0$ and thus reads $\ddot\theta = -\omega_c^2 \,\sin\theta$. The solution derived in Sec.~\ref{ss:EulerLagrange} is inapplicable to this case since we divided Eq.~\eqref{eq:EqofMotion} by $\omega$ to get Eq.~\eqref{eq:Firstintegration}. We can nevertheless follow a similar path as in Sec.~\ref{ss:EulerLagrange} for this simplified equation of motion. The effective potential \eqref{eq:Veff} with $\omega=0$ gives $V_\mathrm{eff} = V_g(\theta)$ and displays a single minimum at $\theta = 0$ (dashed curve in Fig.~\ref{fig:Veff}). The equation for $\dot\theta(t)$ with $\dot\theta_0 = 0$ corresponding to Eq.~\eqref{eq:Firstintegration} is
\begin{eqnarray}\label{eqpendulum}
\dot\theta(t) - 2\omega_c^2 \cos\theta(t) = {\cal C},
\end{eqnarray}
with ${\cal C} = \dot\theta^2_0 / 2 - \omega_c^2 \cos\theta_0$. This leads to a phase plot similar to that shown in Fig.~\ref{fig:PhaseDiagrams}a. For initial conditions located inside the separatrix there are two possible states of the bead. For $\theta_0 = 0$ we have $\theta(t)=0$, the equilibrium position. For $\theta_0 \in (0,\pi)$ and considering first the case $\dot\theta_0=0$ the solution obtained by integrating Eq.~\eqref{eqpendulum} (see Sec.~\ref{ss:EulerLagrange}) involves a delta-amplitude Jacobi elliptic function 
\begin{widetext}
\begin{eqnarray}
\theta\left(t\right)=2\arctan\left\{ \gamma_{+}\mathrm{dn}\left[\omega_{c}t\cos\left(\frac{\theta_0}{2}\right)\left|\gamma_{+}\right|,\sqrt{\dfrac{\gamma_{+}^{2}-\gamma_{-}^{2}}{\gamma_{+}^{2}}}\right]\right\},
\end{eqnarray}
\end{widetext}
with
\begin{equation}
\gamma_{+}^{2}=\dfrac{1-\cos\theta_{0}}{1+\cos\theta_{0}},\quad\gamma_{-}^{2}=-1.
\end{equation}
This expression is a special case of the general solution \eqref{eq:Solution} and has been discussed in Refs.~\onlinecite{Whittaker37,Lima10,Bowman53}.

The separatrix is obtained from Eq.~\eqref{eqpendulum} and $\dot\theta_0=0$ with $\theta_0=\pi$. Integrating that expression one more time leads to the motion of the bead along the separatrix
\begin{equation}\label{thetat2ndseparatrix}
\theta\left(t\right)=2\arctan\left\{ \sinh\left[\omega_c t+\mathrm{arcsinh}\left(\tan\dfrac{\theta_{0}}{2}\right)\right]\right\}.
\end{equation}
Finally, the free spinning case can be obtained in ways similar to the previous sections. We assume $(\theta_0=\pi,\dot\theta_0\neq 0)$ and obtain Eq.~\eqref{eq:FreeMotion} with the new roots (see also Ref.~\onlinecite{Lima10})
\begin{equation}\label{eq:PendulumRoots}
\gamma_{+}^{2}=-1,\quad\gamma_{-}^{2}=\dfrac{\tfrac{1}{4}\dot{\theta}_{0}^{2}+\omega_c^2}{\tfrac{1}{4}\dot{\theta}_{0}^{2}}.
\end{equation}

\section{Conclusion}\label{s:conclusion}

We introduced Jacobi elliptic functions using integral inversion and discussed their properties and the meaning of their arguments $(u,k)$. The first argument of the JEFs, $u$, was shown to be related to an arc length in elliptic geometry, while the second argument $k$ is the eccentricity. As an application of JEFs, we discussed the motion of the bead on the hoop and derived the complete analytical solution of the problem. Our results reduce to the solution of the pendulum when the hoop rotation frequency vanishes. Interestingly, all motions of the bead on the hoop are written in terms of a Jacobi elliptic function, except for equilibria positions and separatrices, where JEFs reduce to trigonometric and hyperbolic functions or constants. The knowledge of JEFs allows expressing the complete solution of the bead on the hoop problem in an elegant and concise form. These functions and their salient features should be remembered as they may appear in number of other problems in physics.

\begin{acknowledgments}
We gratefully acknowledge the support of the National Science Foundation under grant DMR-0907242, the block grant of CNSM and the graduate research fellowship at CSU Long Beach.
\end{acknowledgments}

\appendix

\section{Additional properties of Jacobi elliptic functions}\label{a:JEFrelations}

We introduced Jacobi elliptic functions in Sec.~\ref{s:Jacobi} and presented their main properties. We summarize in this  Appendix a few additional properties of JEFs that were used in solving the bead on the hoop problem.

The JEFs can be related to each other by an appropriate change of variables. For example, starting with the equation $u=F(y,k)$, its inversion gives $y=\mathrm{sn}(u,k)$. On the other hand, multiplying both sides of the equation by $k$ and changing variable $t'=k\,t$ in $F(y,k)$, we can again invert the equation to obtain $k\,y = \mathrm{sn}(k\,u,k^{-1})$. Thus, we found the first of the following relations
\begin{eqnarray}
\label{JEFcorrespondence1}
k\, \mathrm{sn}(u,k) = \mathrm{sn}(k\,u,k^{-1}),\\
\label{JEFcorrespondence2}
\mathrm{dn}(u,k) = \mathrm{cn}(k\,u,k^{-1}).
\end{eqnarray}
The second relation is obtained from the first, using Eqs.~\eqref{JEFconditions1} and \eqref{JEFconditions2}. We use the second equation in Sec.~\ref{s:beadhoop}. Note that in other references, $k^2=m$ is used.

The derivatives of the JEFs with respect to $u$ can be determined by using their definitions as well. For example, from Eq.~\eqref{snuk} one obtains $d\,\mathrm{sn}u/du =  \mathrm{cn}u\,\mathrm{dn}u$. From the Taylor expansions of the trigonometric functions one can also expand the JEFs in powers of $u$. Addition formulae can be derived from Eqs.~\eqref{snuk}--\eqref{dnuk}. For example,
\begin{eqnarray}\label{addsndn}
\label{addsn}
\mathrm{sn}(u+v) &=& \frac{\mathrm{sn}u\,\mathrm{cn}v\,\mathrm{dn}v + \mathrm{sn}v\, \mathrm{cn}u\,\mathrm{dn}u}{\Delta(u,v)},\\
\label{adddn}
\mathrm{dn}(u+v) &=& \frac{\mathrm{dn}u\,\mathrm{dn}v - k^2\,\mathrm{sn}u\,\mathrm{sn}v\, \mathrm{cn}u\,\mathrm{cn}v}{\Delta(u,v)},
\end{eqnarray}
with $\Delta(u,v) = 1 - k^2\,\mathrm{sn}^2u\, \mathrm{sn}^2v$. This definition relates to $\Delta(u)$ introduced in the paragraph preceding Eq.~\eqref{dnuk} via the equality $\Delta(u)=\Delta(u,K)$.  These addition formulae allow establishing relations between different JEFs as well. We used one of them in Sec.~\ref{ss:EulerLagrange}, namely $k'\,\mathrm{nd}(u,k) = \mathrm{dn}(u+K,k)$. This relation can be obtained using Eq.~\eqref{adddn} with $v=K$.

Finally, the JEFs were defined for $u\in\mathbb{R}$, but their definition can be extended to the complex plane with the help of the above addition formulae. For example, using Eq.~\eqref{addsn} one can decompose $\mathrm{sn}(x+iy)$ into real and imaginary parts. In the complex plane, the JEFs are double-periodic functions with periods given by integer multiples of the complete elliptic integral of the first kind.

Further definitions, properties, and expressions can be found in Refs.~\onlinecite{Bowman53}, \onlinecite{Armitage06,Neville71,Chandrasekharan85,bakerMS12}, and \onlinecite{Byrd71}.



\begin{thebibliography}{99}

\bibitem{Hart49}I.V.~Hart, {\it James Watt and the History of Steam Power} (New York: H. Schuman, 1949) p. 103.

\bibitem{Dickinson67}H.W.~Dickinson, {\it James Watt: Craftsman \& Engineer} (New York: Augustus M. Kelley Publishers, 1967) p. 153.

\bibitem{Goldstein02}H.~Goldstein , C.~Poole, J.~Safko, {\it Classical Mechanics}, 3rd ed. (Addison Wesley, San Francisco, 2002) pp. 28-29, 71, 272.

\bibitem{Taylor05} J.R.~Taylor, {\it Classical Mechanics} (University Science, Sausalito, CA, 2005) pp. 260-264.

\bibitem{Bowman53}F.~Bowman, {\it Introduction to Elliptic Functions}  (Wiley, New York, 1953) pp. 7-15, 26-29.

\bibitem{Lima10}F.M.S.~Lima, ``An accurate formula for the period of a simple pendulum oscillating beyond the small angle regime," Am.~J.~Phys.~{\bf 78}, 1146 (2010). C.G.~Carvalhaes and P. Suppes, ``Approximations for the period of the simple pendulum based on arithmetic-geometric mean", Am.~J.~Phys.~{\bf 76}, 1150 (2008).

\bibitem{Whittaker37}Sir E. T. Whittaker, {\it Analytical Dynamics of Particles}, 4th ed. (Cambridge University Press, United States, 1937), pp. 71-74.

\bibitem{Erdos00} P.~Erd\"os, ``Spiraling the Earth with C. G. J. Jacobi," Am.~J.~Phys.~{\bf 68}, 888-895 (2000).

\bibitem{Fu01} Z.~Fu, ``Jacobi elliptic function expansion method and periodic wave solutions of nonlinear wave equations," Phys.~Lett.~A {\bf 289}, 69-74 (2001).

\bibitem{Kinnersley}W.~Kinnersley, ``Exact large amplitude capillary waves on sheets of fluid," J.~Fluid Mech.~{\bf 77}, 229-241 (1976).

\bibitem{Ochoa06} F.~Ochoa and J.~Clavijo, ``Bead, hoop and spring as a classical spontaneous symmetry breaking problem," Eur.~J.~Phys.~{\bf 27}, 1277 (2006).

\bibitem{Prasolov97} V.~Prasolov and Y.~Solovyev, {\it Elliptic Functions and Elliptic Integrals} (American Mathematical Society, Providence, RI, 1997) pp. 25-26, 40-48, 55, 67-69.

\bibitem{Baker1906} A.~Baker, {\it Elliptic Functions} (Wiley, New York, 1906) pp. 1-14.

\bibitem{Armitage06} J.V.~Armitage and W.F.~Eberlein, {\it Elliptic Functions} (University Press, Cambridge, 2006) pp. 1-50, 232-239.

\bibitem{Neville71}E.~Neville, {\it Elliptic Functions} (Pergamon Press, New York, 1971) pp. 89-113.

\bibitem{Chandrasekharan85}K.~Chandrasekharan, {\it Elliptic Functions} (Springer-Verlag, Berlin, 1985) pp. 1-8.

\bibitem{bakerMS12} T.E.~Baker, Master's thesis, California State University Long Beach (2012).

\bibitem{Spivak06}M.~Spivak. {\it Calculus}, 3rd ed.~(Cambridge University Press, Cambridge, 2006) pp. 382-383.

\bibitem{Byrd71}P.F.~Byrd and M.D.~Friedman, {\it Handbook of Elliptic Integrals for Engineers and Scientists}, 2nd ed.~(Springer-Verlag, Berlin, 1971) pp. 50, 55-56.

\bibitem{Rousseaux09}G.~Rousseaux, ``On the `bead, hoop and spring' (BHS) dynamical system," Nonlin.~Dyn.~{\bf 56}, 315-323 (2009)
\end{thebibliography}
\end{document}